\documentclass[reprint,showpacs,amsmath,amssymb,aps,pra,floatfix]{revtex4-1}

\usepackage{graphicx}% Include figure files
\usepackage{bm}% bold math
\usepackage{braket}
\usepackage{hyperref}% add hypertext capabilities
\usepackage[caption=false]{subfig}

\newcommand\abs[1]{\left|#1\right|}
\newcommand\numberthis{\addtocounter{equation}{1}\tag{\theequation}}

\DeclareMathOperator{\Tr}{Tr}

\begin{document}

\title{Bell inequality tests using asymmetric entangled coherent states in asymmetric lossy environments
}

\author{Chae-Yeun Park}
\author{Hyunseok Jeong}\email{h.jeong37@gmail.com}
\affiliation{Center for Macroscopic Quantum Control, Department of Physics and Astronomy,\\
	Seoul National University, Seoul, 151-742, Korea
}

\date{\today}

\begin{abstract}
We study an asymmetric form of two-mode entangled coherent state (ECS), where the two local amplitudes have different values, for testing the Bell-Clauser-Horne-Shimony-Holt  (Bell-CHSH) inequality. We find that the asymmetric ECSs 
have obvious advantages over the symmetric form of ECSs in testing the Bell-CHSH inequality.  
We further study an asymmetric strategy in distributing an ECS over a lossy environment and find that 
such a scheme can significantly increase violation of the inequality.
\end{abstract}

                            
\maketitle

\section{Introduction}
Entangled coherent states (ECSs) in free-traveling fields \cite{yurke86,mecozzi87,sanders92}
have been found to be useful for various applications such as Bell inequality tests \cite{mann95,filip01,wilson02,jeong03,wenger03,JR2006,stobinska07,jeong08,lee09,gerry09,jeong09,lim12,birby13,birby14}, tests for non-local realism \cite{lg1,lg2}, quantum teleportation \cite{jeong02,wang01,enk01,jeong01,an03}, quantum computation \cite{conchrane99,jeong02_comp,ralph03,lund08,marek10,myers11,kim10}, precision measurements \cite{gerry01,gerry02,campos03,munro02,joo11,hirota11,joo12,zhang13}, quantum repeater \cite{repeater} and quantum key distribution \cite{sergienko14}. 
The ECSs can be realized in various systems that can be described as harmonic oscillators and numerous schemes for their implementing have been suggested \cite{yurke86,yurke87,sanders92,mecozzi87,tombesi87,sanders99,sanders00,gerry97}.
An ECS  in a free-traveling field was experimentally generated using the photon subtraction technique on two approximate superpositions of coherent states (SCSs) \cite{ourjoumtsev09}. A proof-of-principle demonstration of quantum teleportation using an ECS as a quantum channel was performed \cite{nielsen13}.

So far, most of the studies on ECSs have considered symmetric types of two-mode states where the
local amplitudes have the same value.
Since ECSs are generally sensitive to decoherence due to photon losses in testing Bell-type inequalities \cite{wilson02}, it would be worth investigating the possibility of using an asymmetric type of ECS to reduce decoherence effects. In fact, asymmetric ECSs can be used to efficiently teleport  an SCS \cite{ralph03,nielsen13}
and to remotely generate symmetric ECSs  \cite{lund13} in a lossy environment.
In addition to the asymmetry of the ECSs, as a closely related issue, it would be beneficial to study  strategies for distributing the ECSs over asymmetrically lossy environments.
In this paper, we investigate asymmetric ECSs as well as asymmetric entanglement distribution strategies, and find their evident advantages over symmetric ones
for testing the Bell-Clauser-Horne-Shimony-Holt (Bell-CHSH) inequality 
 \cite{clauser69} under various conditions.

The remainder of the paper is organized as follows. In Sec.~\ref{sec:bell_test_prefect}, we discuss the Bell-CHSH inequality tests using asymmetric ECSs with ideal detectors. We consider photon on-off detection and photon number parity detection, respectively, for the Bell-CHSH inequality tests. Section~\ref{sec:bell_test_deco} is devoted to the study of decoherence effects on Bell inequality violations with an asymmetric strategy to share ECSs. We then investigate effects of inefficient detectors in Sec.~\ref{sec:ineff}, and conclude with final remarks in Sec.~\ref{sec:remarks}.

\section{\label{sec:bell_test_prefect}Bell inequality tests with asymmetric entangled coherent states}

In this paper, we are interested in a particular form of two-mode ECSs
\begin{align}
	\ket{{\rm ECS}^\pm} = \mathcal{N}_\pm \left( \ket{\alpha_1}\otimes\ket{\alpha_2} \pm \ket{-\alpha_1}\otimes\ket{-\alpha_2} \right) \label{ECS}
\end{align}
where $\ket{\pm\alpha_i}$ are coherent states of amplitudes $\pm\alpha_i$ for field mode $i$, and 
$\mathcal{N}_\pm = [2 \pm 2 \exp ( -2\alpha_1^2-2\alpha_2^2 )]^{-1/2}$ are the normalization factors.  We note that amplitudes $\alpha_1$ and $\alpha_2$ are assumed to be real  without loss of generality throughout the paper.
The ECSs show noticeable properties as macroscopic entanglement when the amplitudes are sufficiently large and
these properties have been extensively explored \cite{JR2006,jeong14,lim12,jeong-review}. It is worth noting that there are studies on other types of ECSs such as multi-dimensional ECSs \cite{EnkInfinite}, cluster-type ECSs  \cite{Semiao1,AnCluster,Semiao2}, multi-mode ECSs \cite{W-ECS} and generalization of ECSs with thermal-state components \cite{JR2006,Pater2006} while we focus on two-mode ECSs in free-traveling fields in this paper.

We call $\ket{{\rm ECS}^+}$ ($\ket{{\rm ECS}^-}$) the even (odd) ECS  because it contains only even (odd) numbers of photons. 
When $\alpha_1=\alpha_2$, we call the states in Eq.~(\ref{ECS}) symmetric ECSs, and otherwise they shall be called asymmetric ECSs.
It is straightforward to show that an ECS can be generated by passing an SCS in the form of $|\alpha\rangle\pm|-\alpha\rangle$ \cite{Ourj2007}, where $\alpha^2=\alpha_1^2+\alpha_2^2$, through a beam splitter. A beam splitter with an appropriate ratio should be used to generate an asymmetric ECS with desired values of $\alpha_1$ and $\alpha_2$.
In this paper, asymmetric ECSs with various values of $\alpha_1$ and $\alpha_2$ are compared for given values of $\alpha$.
The average  photon numbers of the ECSs are solely dependent on the values of $\alpha$ and are
simply obtained as
\begin{align}
\bar{n}_\pm=\braket{{\rm ECS}^\pm|\hat{n}|{\rm ECS}^\pm}
 =  \frac{\alpha^2 (1\mp \,\mathrm{e}^{ -2 \alpha^2})}
{1\pm\, \mathrm{e}^{-2\alpha^2}} \label{number}
\end{align}
where $\hat{n}=\hat{n}_1+\hat{n}_2$ and $\hat{n}_i$ is the number operator for field mode $i$.

\subsection{Bell-CHSH tests with photon on-off measurements}

We first investigate a Bell-CHSH inequality test using photon on-off measurements with the displacement operations. 
A Bell-CHSH inequality test requires parametrized measurement settings of which the outcomes are dichotomized to be either $+1$ or $-1$ \cite{clauser69}. In the simplest example of a two-qubit system, a parametrized rotation about the $x$ axis followed by a dichotomized measurement in the $z$ direction is used. 
In our study, the displacement operator that is known to well approximate the qubit rotation for a coherent-state qubit  \cite{jeong02_comp,jeong03} is used for parametrization. The displacement operator can be implemented using a strong coherent field and a beam splitter with a high transmissivity.
Together with the displacement operator, the photon on-off measurement is experimentally available using current technology with an avalanche photodetector \cite{takeuchi99} in spite of the issue of the detection efficiency. We shall analyze the effects of the limited detection efficiency in Sec. IV.

The photon on-off measurement operator for mode $i$ is defined as
\begin{align}
\hat{O}_i(\xi) = \hat{D}_i(\xi)\left( \sum \limits_{n=1}^{\infty} \ket{n}\bra{n}-\ket{0}\bra{0}\right)\hat{D}_i^\dagger(\xi)
\label{eq:O}
\end{align}
where $\hat{D}_i(\xi)=\exp \left[ \xi\hat{a}_i^\dagger - \xi^*\hat{a}_i\right]$ is the displacement operator and $|n\rangle$ denotes the Fock state.
The correlation function is defined as the expectation value of the joint measurement
\begin{align}
	E_O(\xi_1,\xi_2) = \langle \hat{O}_1(\xi_1)\otimes\hat{O}_2(\xi_2) \rangle \label{E_onoff}
\end{align} 
and the Bell-CHSH function is
\begin{align}
	\mathcal{B}_O = E_O(\xi_1,\xi_2) + E_O(\xi_1',\xi_2)+E_O(\xi_1,\xi_2')-E_O(\xi_1',\xi_2'). \label{bell_onoff}
\end{align}
In any local realistic theory, the absolute value of the Bell-CHSH function is bounded by 2 \cite{clauser69}.

We calculate an explicit form of the correlation function $E_O(\xi_1,\xi_2)$
using Eqs.~(\ref{eq:O}) and (\ref{E_onoff}), of which the details are presented in 
in Appendix \ref{sec:corr_func}.
We then find the Bell-CHSH function $\mathcal{B}_O$ using Eq.~(\ref{bell_onoff}) and its absolute maximum values $\abs{\mathcal{B}_O}_{\rm max}$ over the displacement variables $\xi_1$, $\xi_1'$, $\xi_2$ and $\xi_2'$ together with
$\bar{n}_\pm$, $\alpha_1$ and $\alpha_2$. It requires a numerical multivariable maximization method, and we use the Broyden-Fletcher-Goldfarb-Shanno (BFGS) algorithm \cite{fletcher87} throughout this paper. 

\begin{figure}[t]
	\huge
	\centering

	\subfloat{\resizebox{0.35\textwidth}{!}{
		\put(340,210){\makebox(0,0)[r]{\strut{}\Huge (a)}}
		\includegraphics{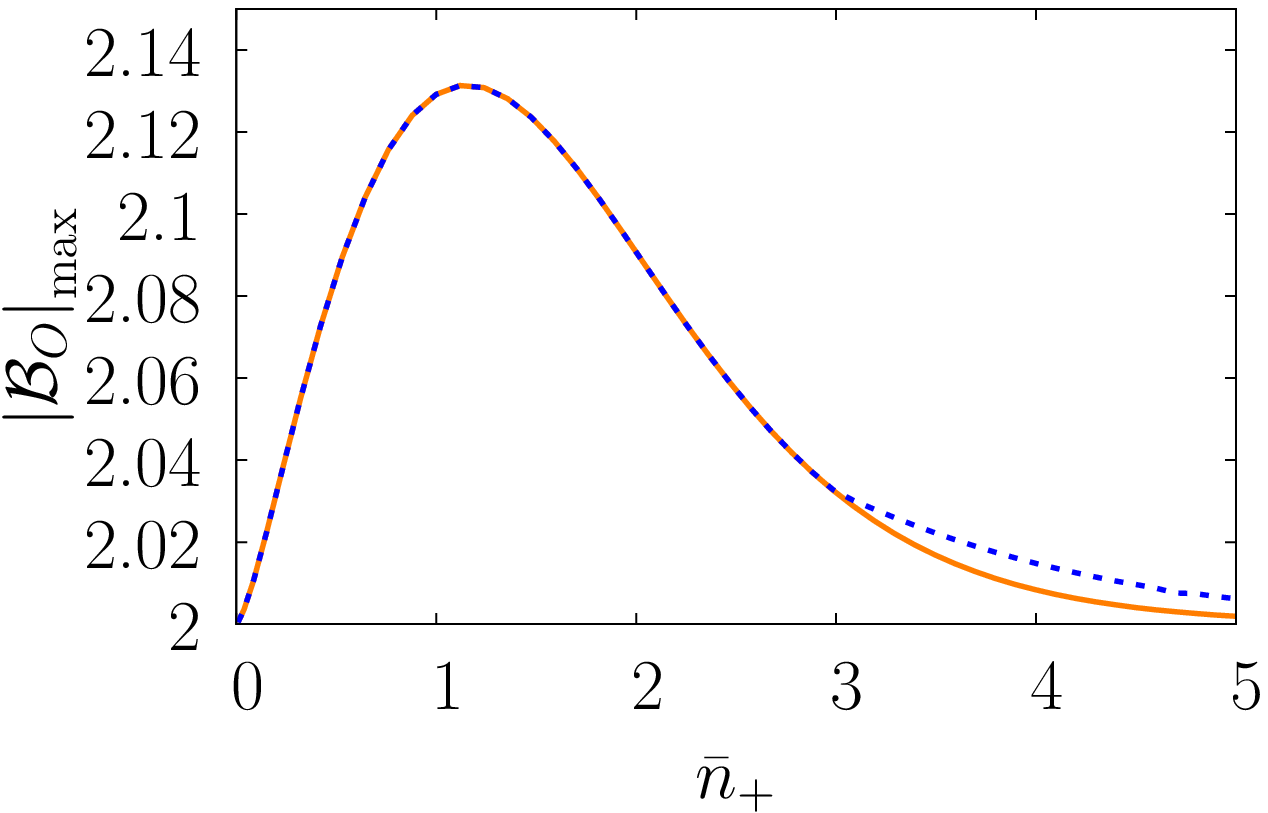}
	}}\\
	\subfloat{\resizebox{0.35\textwidth}{!}{
		\put(340,200){\makebox(0,0)[r]{\strut{}\Huge (b)}}
		\includegraphics{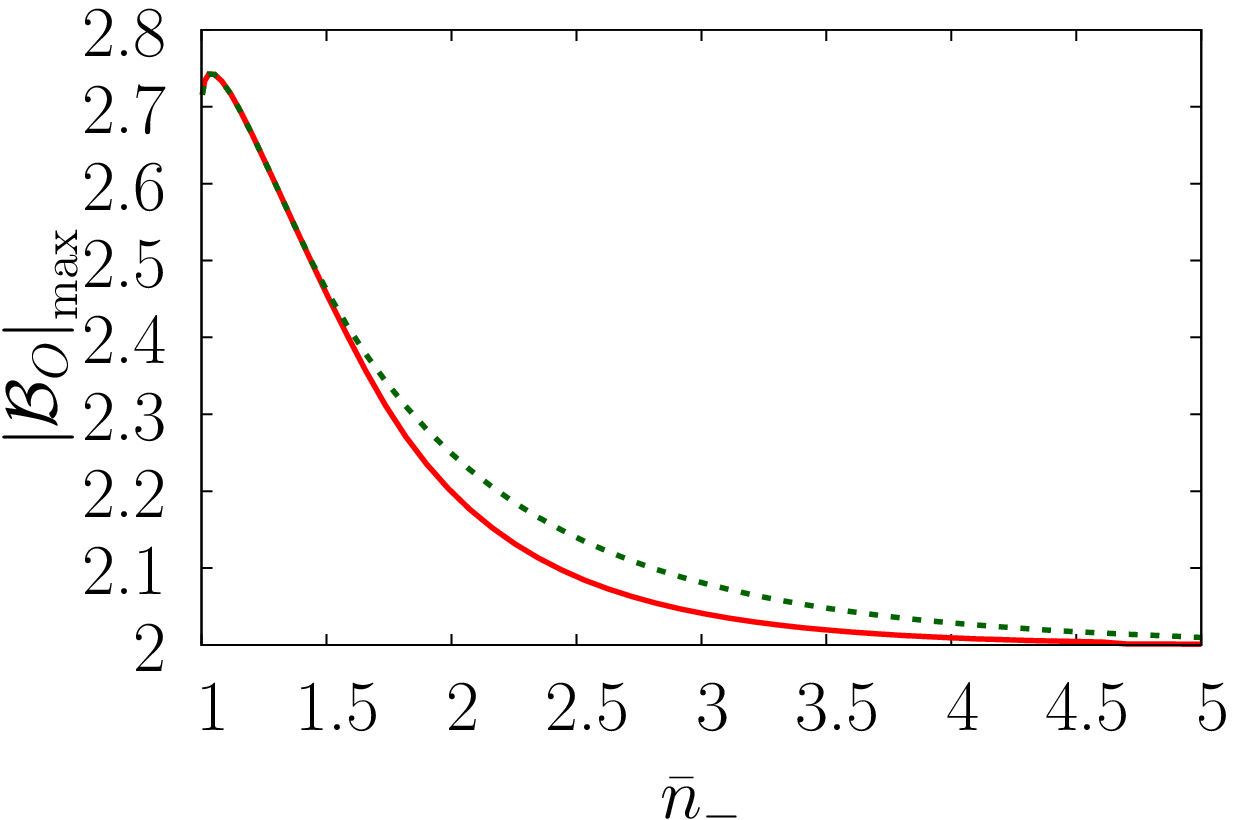}
	}}

	\caption{\label{onoff_n} (Color online) Optimized Bell-CHSH function against the average photon number for (a)
$|{\rm ECS}^+\rangle$ and (b) $|{\rm ECS}^-\rangle$ using photon on-off measurements. The solid curves indicate Bell-CHSH functions optimized over the displacement variables $\xi_1$, $\xi_1'$, $\xi_2$ and $\xi_2'$  for the symmetric ECSs ($\alpha_1=\alpha_2=\alpha/\sqrt{2}$). 
The dotted curves indicate the optimized Bell-CHSH functions with asymmetric ECSs. In the latter case, in addition to the displacement variables, amplitudes $\alpha_1$ and $\alpha_2$ are also optimized under the condition Eq.~\eqref{number}.}
\end{figure}

The optimized Bell-CHSH functions against the average photon numbers $\bar{n}_\pm$ for $|{\rm ECS}^\pm\rangle$ are presented in Fig.~\ref{onoff_n}, while Fig.~2 shows how asymmetric the ECSs become in order to maximize the Bell violations.
The solid curves show the results for symmetric ECSs while the dotted curves correspond to general cases (asymmetric ones). The Bell-CHSH functions are numerically optimized over all displacement variables and amplitudes under the condition of Eq.~\eqref{number}. 
The Bell violations occur both for the even and odd ECSs regardless of the values of $\bar{n}_\pm$ that are consistent with the results in Ref.~\cite{jeong03}.
The odd ECS, $|{\rm ECS}^-\rangle$, violates the inequality more than the other one, $|{\rm ECS}^+\rangle$ for a given average photon number. The Bell-CHSH function for state $|{\rm ECS}^-\rangle$ reaches up to $\abs{\mathcal{B}_O}_{\rm max}\approx2.743$ while the maximum Bell-CHSH function for $|{\rm ECS}^+\rangle$ is $\abs{\mathcal{B}_O}_{\rm max}\approx2.131$. This is due to the fact that the odd ECS is maximally entangled regardless of the value of $\bar{n}_{-}$
\cite{non-ortho,jeong01}
   and the on-off measurement can effectively reveal nonlocality of the ECSs when the amplitudes are small \cite{jeong03}. %\cite{jpark12,kwon13}.

\begin{figure}[t]
	\Large
	\centering
	\resizebox{0.38\textwidth}{!}{
		\includegraphics{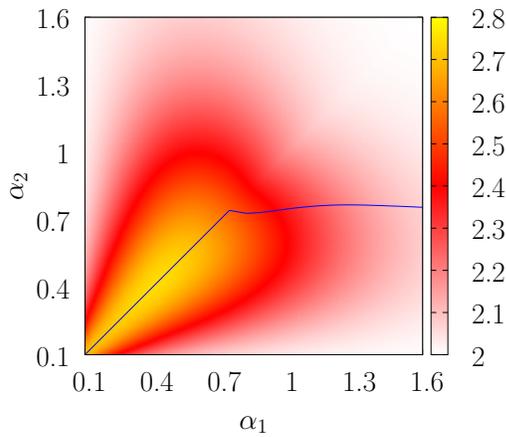}
	}\caption{\label{onoff_minus_ab} (Color online) Optimized Bell-CHSH function $\abs{\mathcal{B}_O}_{\rm max}$ using photon on-off measurements as a function of amplitudes $\alpha_1$ and $\alpha_2$ for $\ket{\rm ECS^-}$. For each point of $\alpha_1$ and $\alpha_2$, the Bell-CHSH function is optimized over the displacement variables. The blue line indicates the states that show the maximum Bell violations among the states for the same the average photon numbers.}
\end{figure}

It is obvious from  Fig.~\ref{onoff_n}  that one can increase the amount of violations
by using the asymmetric ECSs for certain regions of $\bar{n}_{\pm}$.
In the case of state $|{\rm ECS}^-\rangle$,
this difference appears for $\bar{n}_{-}\gtrsim1.43$. This difference in the optimized Bell-CHSH function reaches its maximum value $\approx0.053$ for $\bar{n}_{-}\approx2.24$. 
For this value of $\bar{n}_{-}$, the symmetric ECS gives $|B_O|_\mathrm{max}\approx 2.135$ for $\alpha_1=\alpha_2\approx1.04$ while the asymmetric ECS yields $|B_O|_\mathrm{max}\approx 2.189$ for $\alpha_1\approx1.26$ and $\alpha_2\approx0.77$.  
The even ECS also shows a small increase of the violation when an asymmetric ECS is used in place of the symmetric ECS for $\bar{n}_{+}\gtrsim2.83$. The maximum difference is $\approx0.007$ when $\bar{n}_{+}\approx 3.93$.

In order to further investigate the advantages of the asymmetric ECS, we plot the optimized Bell-CHSH function $\abs{\mathcal{B}_O}_{\rm max}$ as a function of $\alpha_1$ and $\alpha_2$ for $\ket{\rm ECS^-}$ in Fig.~\ref{onoff_minus_ab}. 
The blue line indicates the point for the maximum Bell-CHSH function for each value of $\bar{n}_{-}$. 
The unsmooth change in the blue line at $\alpha_1=\alpha_2\approx 0.77$ ($\bar{n}_-\approx1.43$) results from the numerical optimization process where {\it local} maximum values are compared with changes of related parameters, i.e., the displacement variables and amplitudes.
We note that such comparisons among local maxima at a number of different parameter regions lead to unsmooth changes in several plots throughout this paper \cite{note-num}.
In fact, the blue line splits to two symmetric curves from the point of $\bar{n}_-\approx1.43$, according to our numerical calculation (which is obvious  because $\alpha_1$ and $\alpha_2$ are simply interchangeable), while we plot only one of the curves for convenience.

The results in Figs.~\ref{onoff_n}  and \ref{onoff_minus_ab} shows that when the average photon number is relatively large, the asymmetric ECS outperforms 
the symmetric one in testing the Bell-CHSH inequality with photon  on-off measurements and displacement operations. 
On the other hand, when the average photon number is small, the symmetric ECS gives larger Bell violations.

\begin{figure}[t]
	\huge
	\centering
	\subfloat{\resizebox{0.35\textwidth}{!}{
		\put(100,210){\makebox(0,0)[r]{\strut{}\Huge (a)}}
		\includegraphics{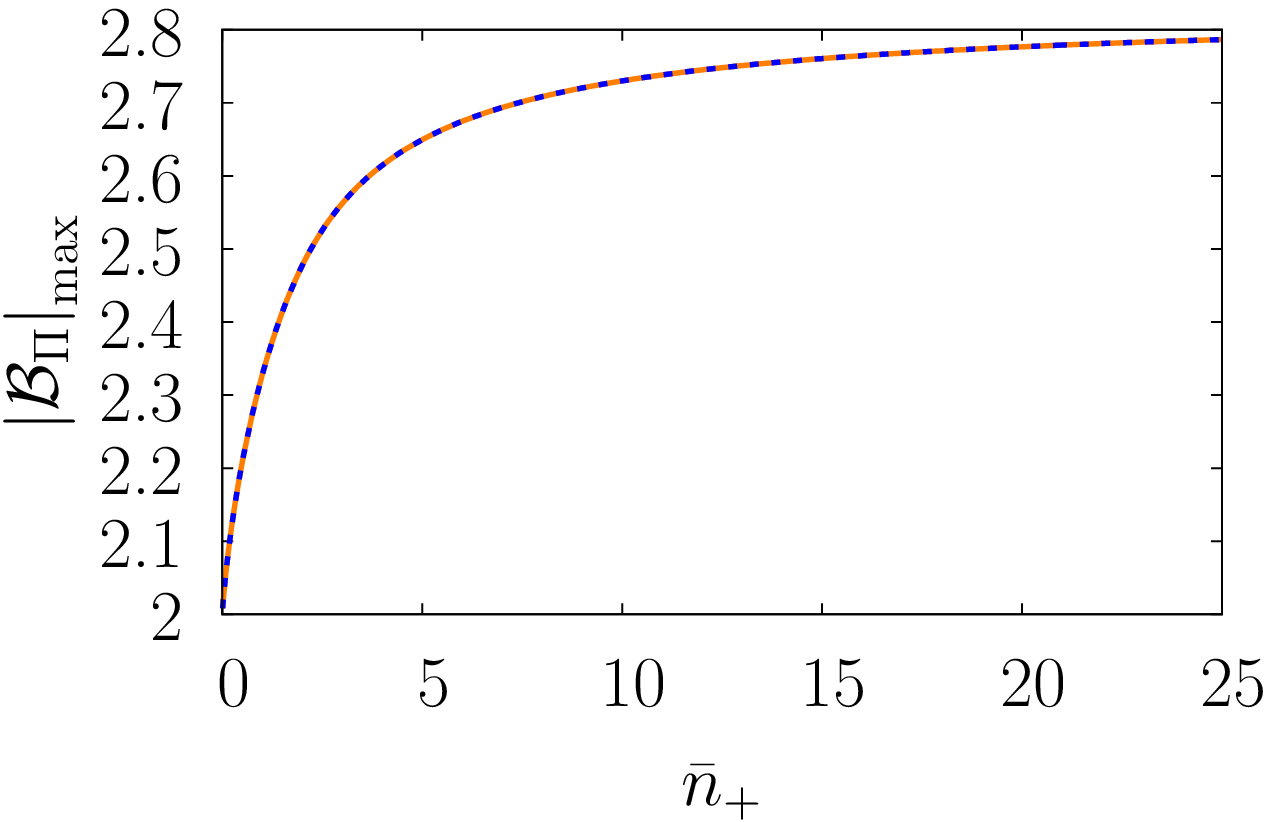}
	}}\\
	\subfloat{\resizebox{0.35\textwidth}{!}{
		\put(100,205){\makebox(0,0)[r]{\strut{}\Huge (b)}}
		\includegraphics{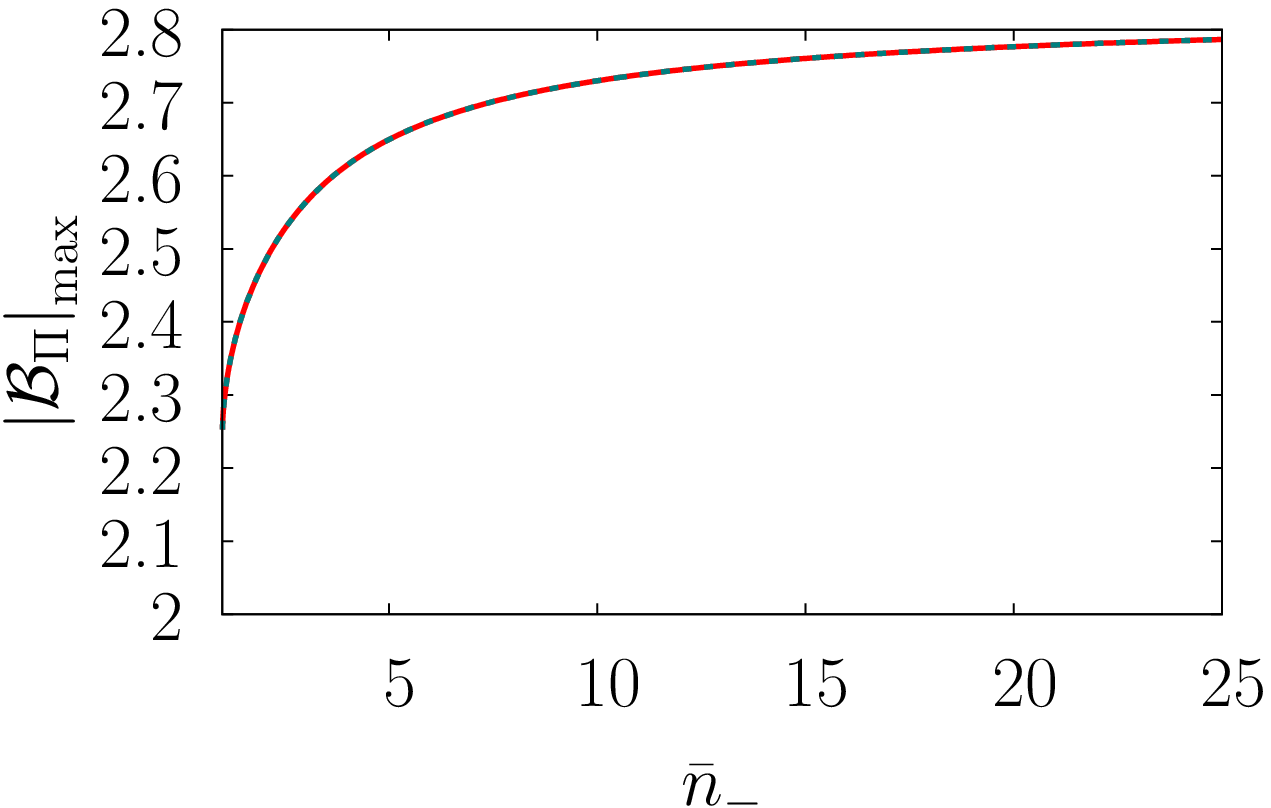}
	}}
	\caption{\label{nvsb_parity} (Color online) Optimized Bell-CHSH functions using photon  number parity measurements against average photon number of (a) the even ECS and (b) the odd ECS. The solid curves show the optimization results for symmetric form of ECS. The dotted curves are the results for general form of ECSs. Like the on-off measurement case, we optimized the displacement variables and amplitudes for each setting. }
\end{figure}

\subsection{Bell-CHSH tests with photon number parity measurements}
We now consider the photon number parity measurements with the displacement operators for both modes. 
Here we use the displacement operator again in order to approximate a parametrized rotation for a coherent-state qubit in the Bell-CHSH test.
The displaced parity measurement is given by
\begin{align}
	\hat{\Pi}_i(\xi)=\hat{D}_i(\xi) \sum \limits_{n=0}^{\infty}\big( \ket{2n}\bra{2n}-\ket{2n+1}\bra{2n+1} \big) \hat{D}^\dagger_i(\xi) \label{eq:Pi}
\end{align}
where $i\in\{1,2\}$ denotes each mode
and the correlation function is 
\begin{align}
	E_\Pi(\xi_1,\xi_2) = \braket{\hat{\Pi}_1(\xi_1)\otimes\hat{\Pi}_2(\xi_2)} \label{E_parity}
\end{align}
with which the Bell-CHSH function $\mathcal{B}_\Pi$ can be constructed using Eq.~\eqref{bell_onoff}.
We have obtained and presented an explicit form of the correlation function in Appendix \ref{sec:corr_func}.

\begin{figure}[t]
	\Large
	\centering
	\resizebox{0.38\textwidth}{!}{
		\includegraphics{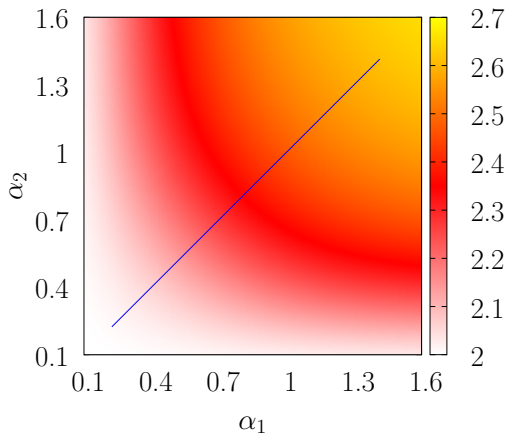}
	}
	\caption{\label{parity_plus_ab} (Color online) Optimized Bell-CHSH function for $\ket{\rm ECS^+}$ using photon number parity measurements.
	We optimize the Bell-CHSH function for amplitudes, $\alpha_1$ and $\alpha_2$, with all displacement variables, and find that the function is optimized when  $\alpha_1=\alpha_2$.}
\end{figure}

The optimized Bell-CHSH functions are plotted for varying $\bar{n}_{\pm}$ using parity measurements in Fig.~\ref{nvsb_parity}. In contrast to the on-off measurement case, the optimized Bell-CHSH functions increase monotonically toward Cirel'son's bound \cite{cirelson80}, $2\sqrt{2}$, for both even and odd ECSs as shown in the figure. These results are consistent with previous studies \cite{wilson02,jeong03}. 
As explained in Ref.~\cite{jeong03}, the reason that the Bell-CHSH violation increases monotonically in the case of the parity measurement but not in the case of the on-off measurement can be explained as follows. When the average photon number of the symmetric ECS is sufficiently large, it can be represented as a maximally entangled two-qubit state in a $2 \otimes 2$ Hilbert space spanned by the even and odd SCS basis ($|\alpha\rangle\pm|-\alpha\rangle$) and the displacement operator well approximates the qubit rotation. Therefore, in this limit, the violation approaches the maximum value, $2\sqrt{2}$, with the parity measurement that perfectly discriminates between the even and odd SCS. In contrast, the photon number on-off measurement cannot cause Bell violations for large amplitudes because the vacuum weight in the state almost vanishes in the limit of $\alpha\gg1$~\cite{jeong03}.

As implied by the apparent overlaps between the cases of symmetric ECSs (solid curve) and those of the asymmetric ECSs (dotted), our numerical investigations show that unlike the case with on-off measurements, the asymmetric ECSs do not show any larger Bell violations. 
Figure~\ref{parity_plus_ab} shows the optimized Bell-CHSH function of $\ket{\rm ECS^+}$ for amplitudes $\alpha_1$ and $\alpha_2$. The blue line in the figure indicates the states with which the maximum violations are obtained for the same average photon numbers. 
The figure shows that the asymmetry of the amplitudes does not improve the amount of violations in the case of the parity measurement. We conjecture that the symmetric structure of the parity measurement is closely related to this result.

\section{\label{sec:bell_test_deco}Bell-CHSH inequality tests under decoherence effects}
In this section, we consider decoherence effects on Bell nonlocality tests using an ECS. We will study the Bell-CHSH inequality violation using a general form of ECS and compare it to the result with a symmetric ECS for both on-off and parity measurements.

\subsection{Symmetric and asymmetric strategies for entanglement distribution}

In the presence of photon loss, the time evolution of a density operator $\rho$ is described by the master equation as \cite{phoenix90}
\begin{align*}
	\frac{\partial\rho}{\partial\tau} = \hat{J}\rho + \hat{L}\rho, \label{master}\numberthis
\end{align*}
where $\tau$ is the interaction time, and the Lindblad superoperators $\hat{J}$ and $\hat{L}$ are defined as $\hat{J}\rho = \gamma \sum_i \hat{a}_i \rho \hat{a}_i^\dagger$ and $\hat{L}\rho = - \gamma/2 \sum_i (\hat{a}_i^\dagger \hat{a}_i \rho + \rho \hat{a}_i^\dagger \hat{a}_i)$ with a decay rate $\gamma$.

\begin{figure}[t]
\centering
	\resizebox{0.48\textwidth}{!}{
		\includegraphics{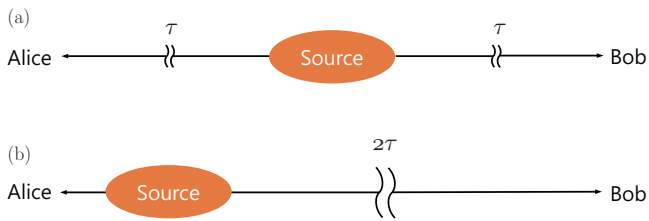}
	}
	\caption{\label{photon_loss} Schemes to distribute an ECS in a symmetric way (a) and in an asymmetric way (b) strategies. (a) In strategy A, photon losses occur in both parts with the same rate and for time $\tau$.  (b) In strategy B, an ECS is generated in the location of Alice and one of the field modes is sent to Bob. Photon losses then occur only one of the two modes for time $2\tau$. }
\end{figure}

We consider two different strategies, i.e. symmetric (A) and asymmetric (B) ones, when distributing an ECS over a distance to Alice and Bob as illustrated in Fig.~\ref{photon_loss}. In the case of strategy A, photon loss occurs symmetrically for both modes of the ECS during the decoherence time $\tau$. On the other hand, in strategy B, an ECS is first generated in the location of Alice and one mode of the ECS is sent to Bob at a distance. In the latter case, photon losses occur only in one of the field modes but the decoherence time becomes $2\tau$. Assuming a zero temperature bath and a decay rate $\gamma$ for both cases, a direct calculation of the master equation leads to the states
\begin{align*}
	\rho^{\pm}_A(\alpha_1,\alpha_2,t)	&= \ \mathcal{N_\pm}^2 \Bigl\{ \ket{\sqrt{t}\alpha_1}\bra{\sqrt{t}\alpha_1}\otimes\ket{\sqrt{t}\alpha_2}\bra{\sqrt{t}\alpha_2} \\
	&\pm e^{-2(1-t)(\alpha_1^2+\alpha_2^2)}&\\ 
	&\big[\ket{\sqrt{t}\alpha_1}\bra{-\sqrt{t}\alpha_1}\otimes\ket{\sqrt{t}\alpha_2}\bra{-\sqrt{t}\alpha_2} \\
	&+ \ket{-\sqrt{t}\alpha_1}\bra{\sqrt{t}\alpha_1}\otimes\ket{-\sqrt{t}\alpha_2}\bra{\sqrt{t}\alpha_2} \big] \\
	&+ \ket{-\sqrt{t}\alpha_1}\bra{-\sqrt{t}\alpha_1}\otimes\ket{-\sqrt{t}\alpha_2}\bra{-\sqrt{t}\alpha_2} \Bigr\}, \numberthis \label{rhoA}\\
	\rho^\pm_B(\alpha_1,\alpha_2,t)
	&= \ \mathcal{N_\pm}^2 \Bigl\{ \ket{\alpha_1}\bra{\alpha_1}\otimes\ket{t\alpha_2}\bra{t\alpha_2} \\
	& \pm e^{-2(1-t^2) \alpha_2^2} \bigl[ \ket{\alpha_1}\bra{-\alpha_1}\otimes\ket{t\alpha_2}\bra{-t\alpha_2} \\
	&+ \ket{-\alpha_1}\bra{\alpha_1}\otimes\ket{-t\alpha_2}\bra{t\alpha_2} \bigr]\\
	&+ \ket{-\alpha_1}\bra{-\alpha_1}\otimes\ket{-t\alpha_2}\bra{-t\alpha_2} \Bigr\} \numberthis \label{rhoB}
\end{align*}
for strategies A and B, respectively, where $t = e^{-\gamma \tau}$. For convenience, we define the normalized time $r=1-t$ which has the value of zero when $\tau=0$ and increases to $1$ as $\tau$ increases to the infinity. 
If the cross terms in Eqs.~\eqref{rhoA} and \eqref{rhoB} vanish, the states become classical mixtures of two distinct states and quantum effects generally disappear. We observe that the cross term in $\rho^{\pm}_A$ is proportional to $e^{-2(1-t)(\alpha_1^2+\alpha_2^2)}$ and it is proportional to $e^{-2(1-t^2) \alpha_2^2}$ in  $\rho^{\pm}_B$. This implies that one may reduce decoherence effects using strategy B by making the amplitude smaller for the mode in which loss occurs (i.e., by making the field mode sent to Bob  in Fig.~\ref{photon_loss} to have the smaller amplitude). Such a strategy was also applied to the tele-amplification protocol \cite{nielsen13} and the distributed generation scheme for ECSs \cite{lund13}.

\begin{figure}[t]
	\huge
	\centering
	\subfloat{\resizebox{0.35\textwidth}{!}{
		\put(110,70){\makebox(0,0)[r]{\strut{}\Huge (a)}}
		\includegraphics{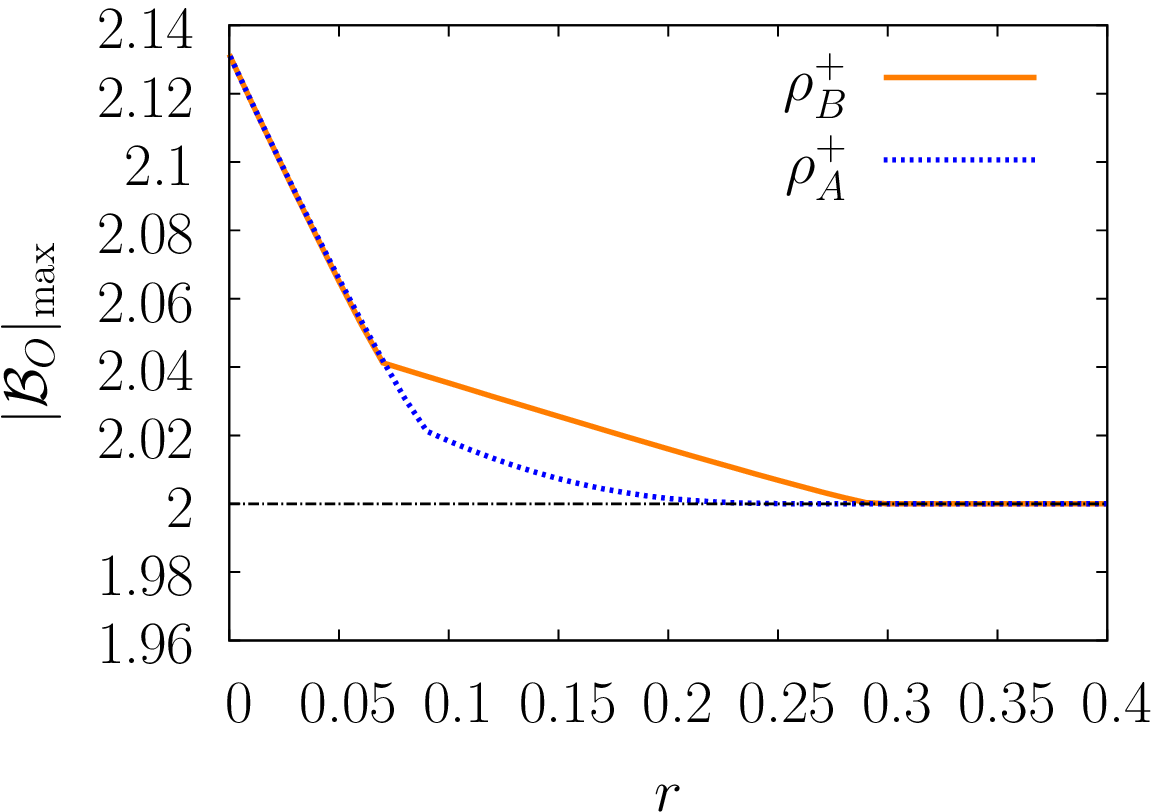}
	}}\\
	\subfloat{\resizebox{0.35\textwidth}{!}{
		\put(105,90){\makebox(0,0)[r]{\strut{}\Huge (b)}}
		\includegraphics{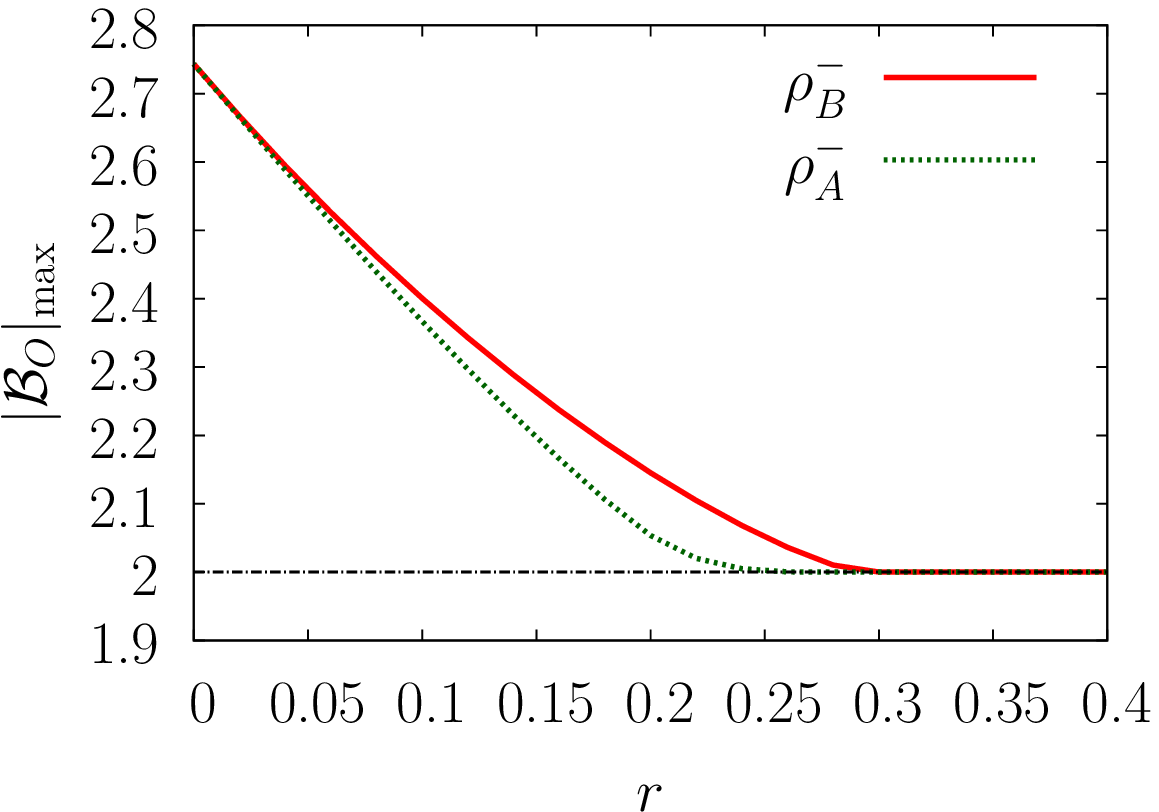}
	}}
	\caption{\label{onoff_t} (Color online) Optimized Bell-CHSH function $|{\cal B}_O|_{\rm max}$ against the normalized time $r$ for (a) even ECSs and (b) odd ECSs using on-off measurements.  The dotted curves indicate the optimized results for strategy A for entanglement distribution and the solid curves correspond to the results for strategy B explained in the main text. Amplitudes $\alpha_1$, $\alpha_2$ together with displacement variables  are all numerically optimized to find the maximum values of the Bell-CHSH function for given $r$. The optimizing values of $\alpha_1$ and $\alpha_2$ are  found between $0.4$ and $1.4$, where relatively smaller values correspond to large values of $r$. The dot-dashed horizontal line indicates the classical limit, $2$.}
\end{figure}

\begin{figure}[t]
	\Large
	\centering
	\subfloat{\resizebox{0.38\textwidth}{!}{
		\llap{\parbox[b]{0in}{\LARGE (a)\\\rule{0ex}{3.3in}}}
		\includegraphics{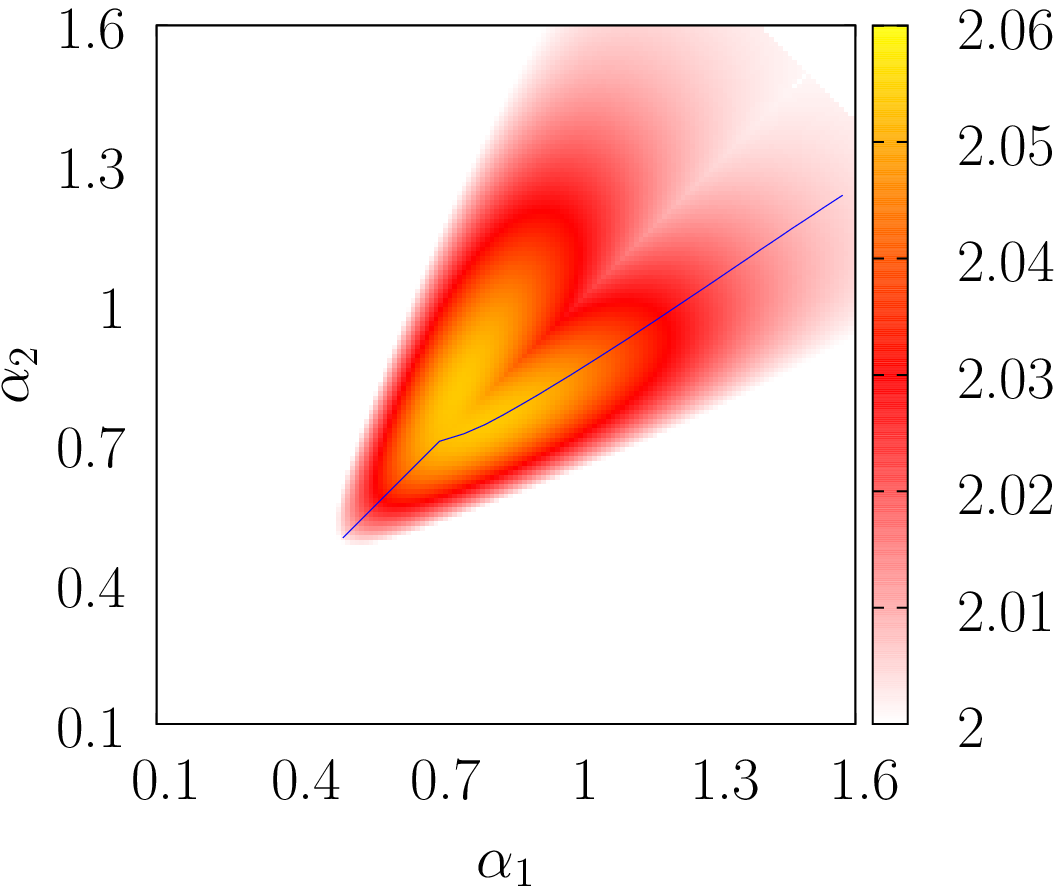}
	}}\\
	\subfloat{\resizebox{0.38\textwidth}{!}{
		\llap{\parbox[b]{0in}{\LARGE (b)\\\rule{0ex}{3.3in}}}
		\includegraphics{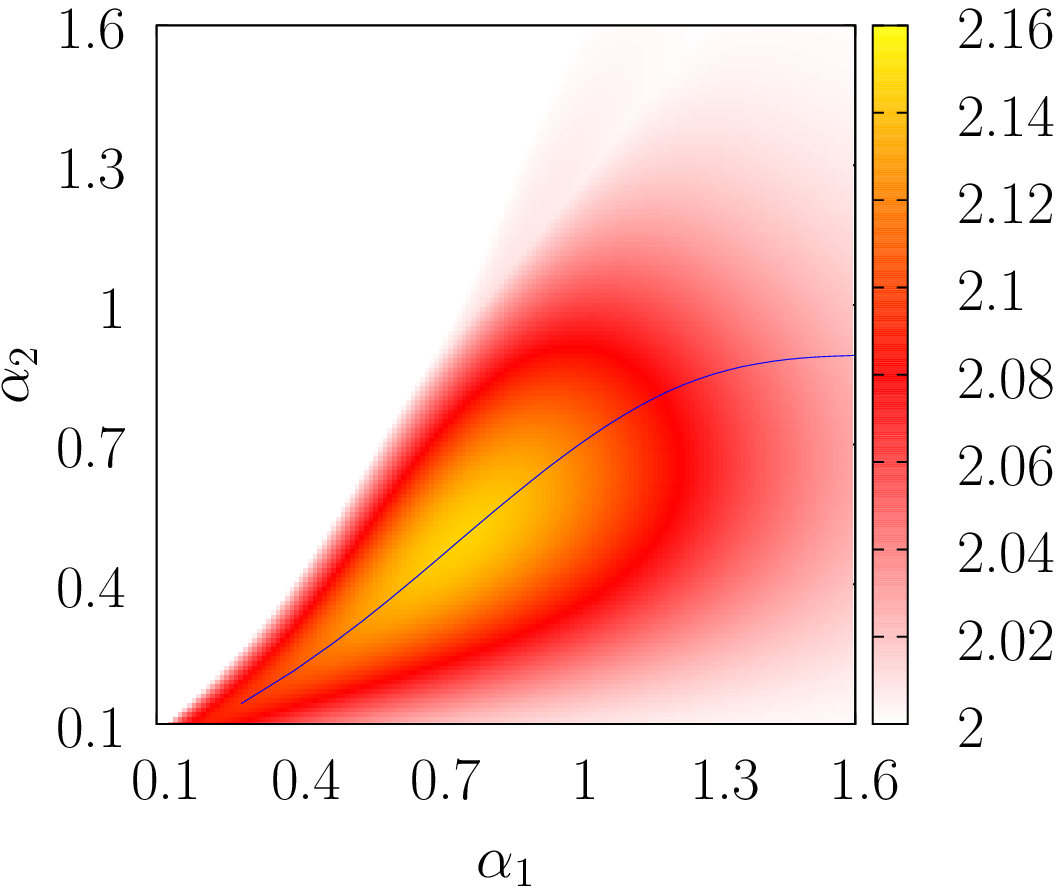}
	}}
	\caption{\label{onoff_ab} (Color online) Optimized Bell-CHSH values against $\alpha_1$ and $\alpha_2$ (a) in strategy A and (b) in strategy B using on-off measurements for the odd ECSs. The normalized time is $r=0.2$ for both (a) and (b). In (a), only one side of the blue curves is displayed. Since the correlation function has a symmetry over the exchange of the two modes, the maximal violation points also exist in the opposite side relative to the center line of $\alpha_1=\alpha_2$. 
	Strategy B generally shows the larger violations of the Bell-CHSH inequality than strategy A. 
The optimizing value of $\alpha_2$ for the same value of $\alpha_1$ with strategy B is always smaller than that with strategy A.}
\end{figure}

\subsection{Bell-CHSH tests with on-off measurements under photon loss effects}

We first consider Bell-CHSH tests with on-off measurements in comparing strategies A and B regarding robustness to the decoherence effects. An explicit form of the correlation functions calculated using Eqs.~\eqref{rhoA} and \eqref{rhoB} can be found Appendix~\ref{sec:corr_func}. We numerically optimize the corresponding Bell-CHSH function, $|{\cal B}_O|_{\rm max}$, over all displacement variables and amplitudes $\alpha_1$ and $\alpha_2$ for given $r$. 
The numerically optimized Bell-CHSH functions against the normalized time are presented for both $\rho_A^\pm$ and $\rho_B^\pm$ in Fig.~\ref{onoff_t} where the decrease of violations due to the decoherence effects is apparent.
The optimizing values of amplitudes $\alpha_1$ and $\alpha_2$ are found between $0.4$ and $1.4$, where relatively smaller values correspond to large values of $r$. We observe that strategy B leads to larger violations than strategy A for $r\gtrsim0.07$ when we use the even ECS $\rho^+$. For the smaller value of $r$, strategy A gives slightly larger violations where the difference is up to $\lesssim 0.001$. On the other hands, the odd ECS $\rho^-$ shows larger violations for strategy B than strategy A regardless of the value of $r$. 
 As explained earlier in Sec.~IIA, the discontinuities of the first derivative at $r\approx0.09$ for $\rho_A^+$ and $r\approx0.07$ for $\rho_B^+$  in Fig.~\ref{onoff_t}(a) emerge from the numerical optimization process where local maxima for different parameter regions are compared.

\begin{figure}[t]
	\huge
	\centering
	\subfloat{\resizebox{0.35\textwidth}{!}{
		\put(100,210){\makebox(0,0)[r]{\strut{}\Huge (a)}}
		\includegraphics{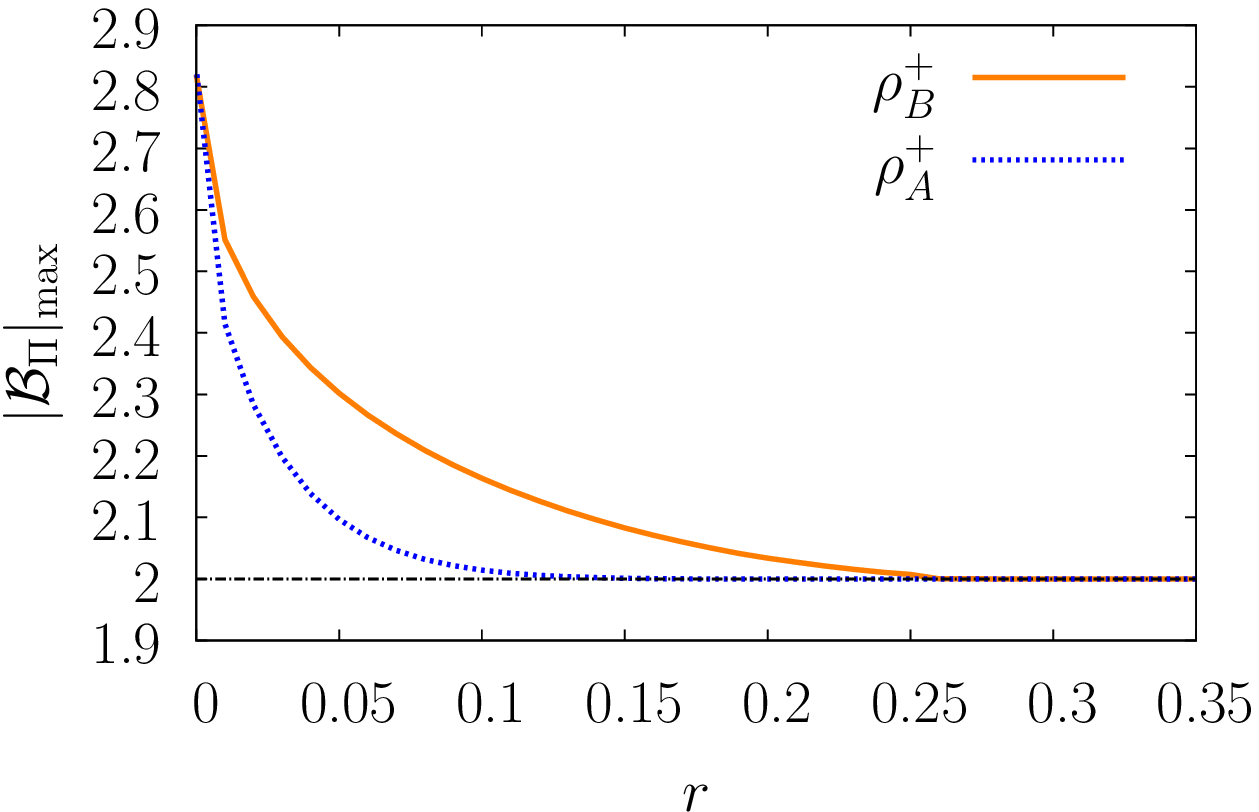}
	}}\\
	\subfloat{\resizebox{0.35\textwidth}{!}{
		\put(100,210){\makebox(0,0)[r]{\strut{}\Huge (b)}}
		\includegraphics{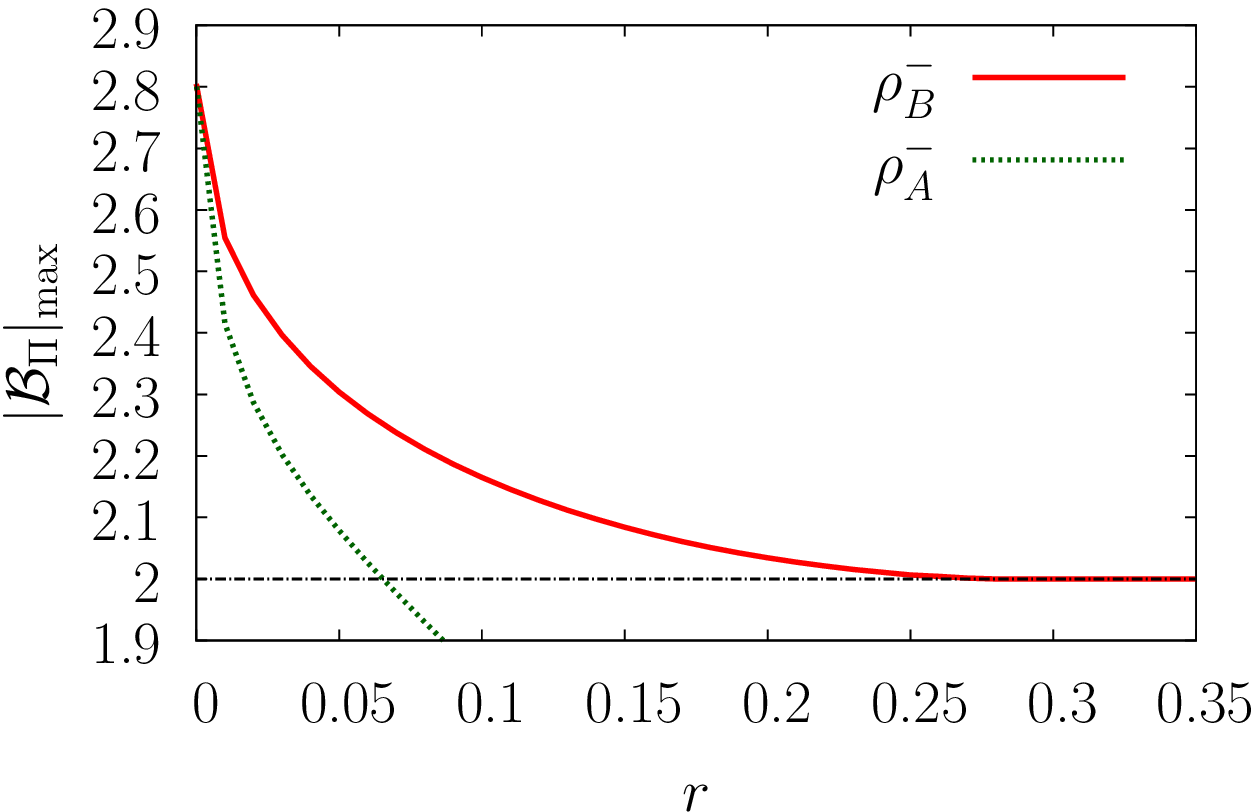}
	}}
	\caption{\label{parity_t} (Color online) Optimized Bell-CHSH function against normalized decoherence time $r$ for photon parity measurement (a) for even ECS and (b) for odd ECS. The dotted curves show the optimized results for strategy A and the solid curves show the results for strategy B. 
The dot-dashed line indicates the classical limit $2$.
}
\end{figure}

The Bell violations of the odd ECSs for varying $\alpha_1$ and $\alpha_2$ for $r=0.2$, numerically optimized
for the displacement variables, are shown in Fig.~\ref{onoff_ab}. Figure~\ref{onoff_ab}(a) clearly shows the asymmetric behavior of the optimized Bell-CHSH function under strategy A.  Figure~\ref{onoff_ab}(b) is for the case of asymmetric decoherence (strategy B). For a given average photon number, the optimizing value of $\alpha_2$ in strategy B is smaller than that of strategy A. It shows that an asymmetric ECS can reduce the asymmetric decoherence effects by lessening the amplitude of the mode which decoherence occurs and adjusting the other mode which is free from decoherence. These results are consistent with previous studies for Bell inequality tests using hybrid entanglement \cite{jpark12,kwon13} and quantum teleportation \cite{kpark12,nielsen13}, where the amplitude of the mode that suffers decoherence should be kept small in order to optimize Bell violations or teleportation fidelities.

It is important to note from Fig.~\ref{onoff_ab} that strategy B shows significantly larger violation than that of strategy A. In strategy A (Fig.~\ref{onoff_ab}(a)), the maximum value of the Bell-CHSH function is $\approx2.054$ for $\alpha_1\approx 0.81$ and $\alpha_2 \approx 0.74$. On the other hands, the maximum value for strategy B (Fig.~\ref{onoff_ab}(b)) is $\approx2.145$ for $\alpha_1 \approx 0.76$ and $\alpha_2 \approx 0.50$. 
Noting that $2.0$ is the maximum value of the Bell function by a local realistic theory, the absolute value of the Bell function with strategy B, $2.145$, is significantly larger than that of strategy A, $2.054$. It is a remarkable advantage of using asymmetric ECSs with the asymmetric distribution scheme in testing the Bell-CHSH inequality.

\subsection{\label{subsec:parity_loss}Bell-CHSH tests with photon number parity measurement under photon loss effects}

\begin{figure}[t]
	\Large
	\centering

	\subfloat{\resizebox{0.38\textwidth}{!}{
		\llap{\parbox[b]{0in}{\LARGE (a)\\\rule{0ex}{3.3in}}}
		\includegraphics{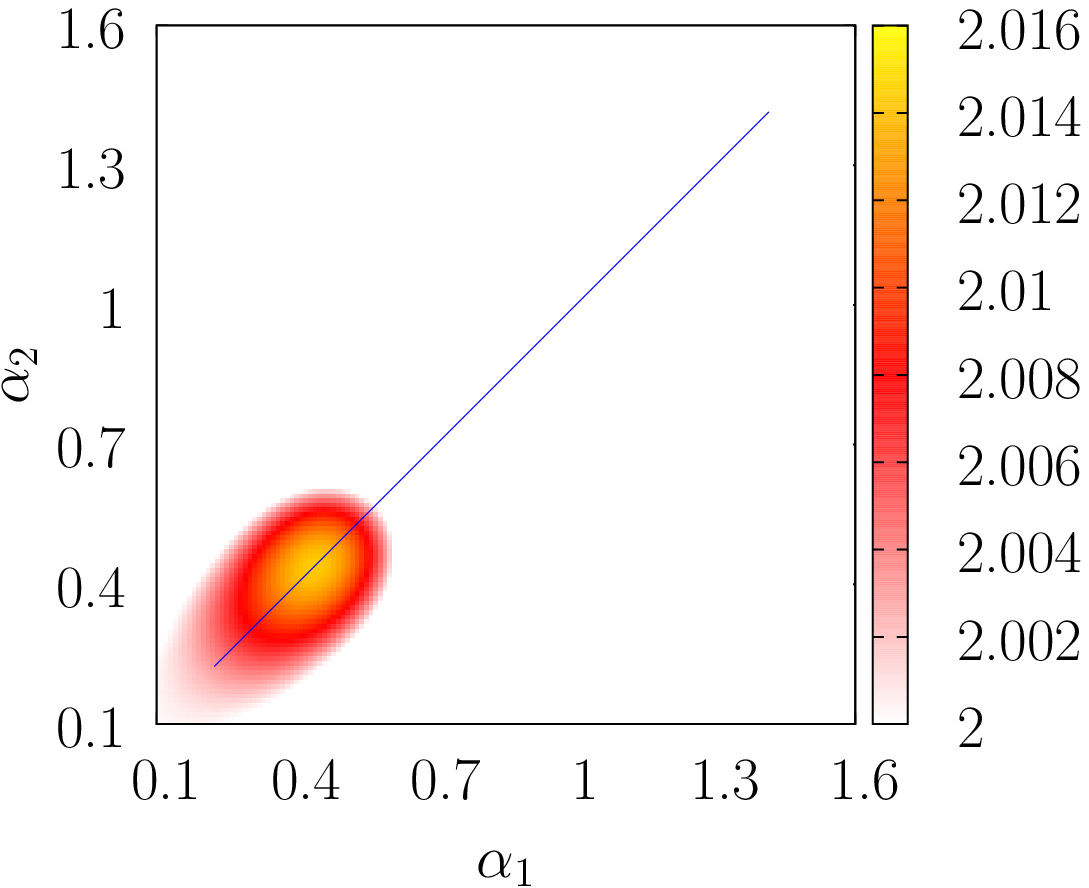}
	}}\\
	\subfloat{\resizebox{0.38\textwidth}{!}{
		\llap{\parbox[b]{0in}{\LARGE (b)\\\rule{0ex}{3.3in}}}
		\includegraphics{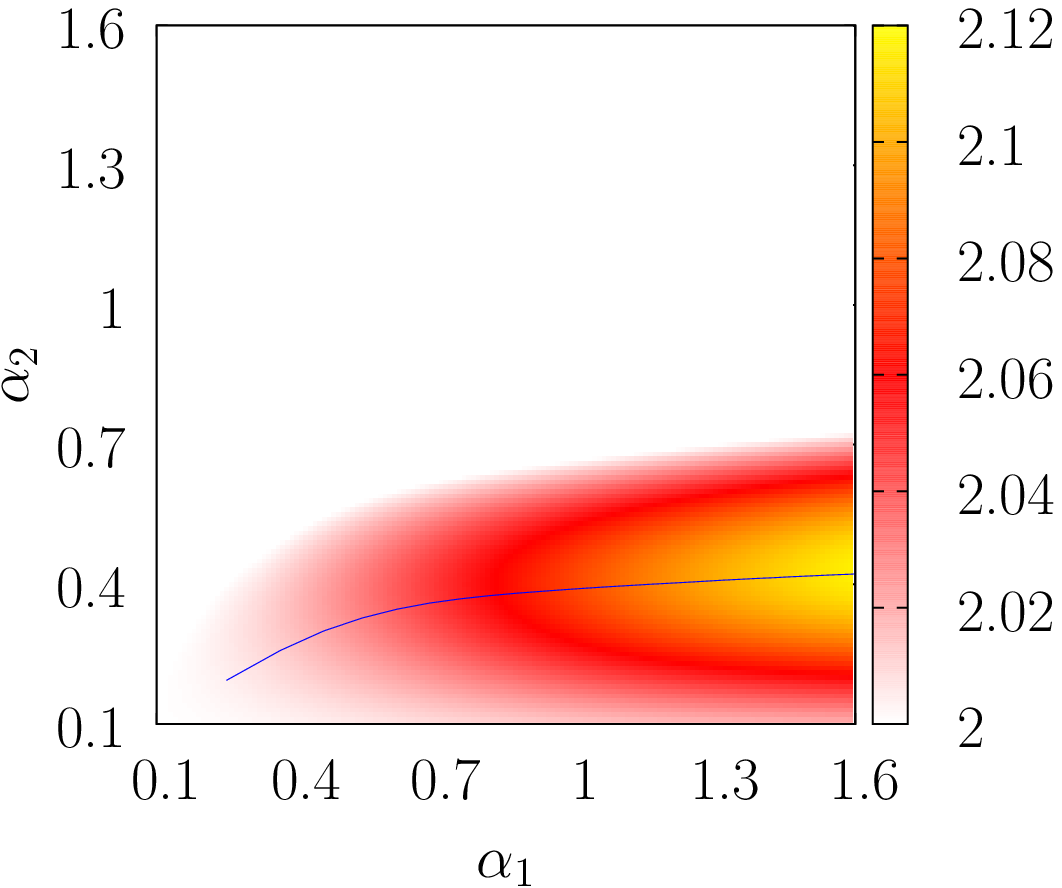}
	}}
	\caption{\label{parity_ab} (Color online) Using parity measurement for even ECSs, we optimize the Bell-CHSH function against $\alpha_1$ and $\alpha_2$ for $r=0.1$ (a) in strategy A and (b) in strategy B. All displacement variables are optimized for given $\alpha_1$ and $\alpha_2$.}
\end{figure}

Figure~\ref{parity_t} presents the numerically optimized Bell-CHSH function $\abs{\mathcal{B}_\Pi}$ with parity measurements against the normalized time $r$  (see Appendix \ref{sec:corr_func} for details). 
For both the even and odd ECSs, strategy B give significantly larger violations over strategy A. The optimizing values of $\alpha_1$ and $\alpha_2$ approach infinity for $r\rightarrow0$. This is because, when we use the photon number parity measurements and the photon loss is absence in one mode, the optimized Bell-CHSH functions increase monotonically with repect to the amplitude of that mode. 
However, we are interested in values that can be obtained within a realistic range (e.g. $\bar{n}_\pm<5$).
We thus plot  in Fig.~\ref{parity_ab} the optimization results for the even ECSs under a normalized decoherence time ($r=0.1$) for varying $\alpha_1$ and $\alpha_2$. In contrast to the case of on-off measurements, as shown in  Fig.~\ref{parity_ab}(a), the asymmetry of the amplitudes of the ECS does not improve the amount of Bell violations for strategy A where decoherence occurs symmetrically. 
Similar to the results of the on-off measurements, the asymmetry of the amplitudes in ECSs increases the amount of Bell violation for strategy B (Fig.~\ref{parity_ab}(b)). In this case, the maximum value of optimized Bell-CHSH function is $|{\cal B}_\Pi|_{\rm max}\approx 2.146$ for $\alpha_1\rightarrow \infty$ and $\alpha_2\approx 0.46$. A similar value $|{\cal B}_\Pi|_{\rm max}\approx 2.131$ can be obtained for $\alpha_1=2$ and $\alpha_2\approx 0.44$ where the average photon number is given by $n_+\approx 4.19$.
This value is much larger than that of  strategy A, which only shows $|{\cal B}_\Pi|_{\rm max}\approx2.014$ for $\alpha_1=\alpha_2 \approx 0.44$.

\section{\label{sec:ineff}Effects of detection inefficiency}

Here we investigate the effects of detection inefficiency in the Bell-CHSH inequality tests.
An imperfect photodectector with efficiency $\eta$ can be modeled using a perfect photodetector together with a beam splitter of transmissivity $\sqrt{\eta}$ in front of it \cite{yuen80}. Meanwhile,  decoherence by photon loss (Eqs.~\eqref{rhoA} and \eqref{rhoB}) in the entangled state can also be modeled using a beam splitter.
The only difference between the effect of detection inefficiency and that of photon loss in the entangled state
is the order of photon loss and displacement as shown in Fig.~\ref{scheme1}. We compare these two cases in Appendix \ref{sec:equiv} where  the results show that the two cases lead the same Bell-CHSH function except for the coefficients of the displacement variables which disappear during the optimization process.

We have numerically investigated both cases of on-off and parity measurements with limited efficiencies, $\eta_1$ for mode 1 and $\eta_2$ for mode 2, for both even and odd ECSs.
As we find that the even ECS is better for on-off measurements and the odd ECS is better for parity measurements, as already implied in Figs. 6 and 8, we present the two corresponding cases in Figure~\ref{inefficient}.
It shows that the on-off measurement scheme generally gives much larger violations compared to the parity measurement scheme. This is consistent with the results of the case with photon loss effects studied in the previous section (see Figs. 6(b) and 8(a)). 

\begin{figure}[t]
	\Large
	\centering
	\resizebox{0.47\textwidth}{!}{
		\includegraphics{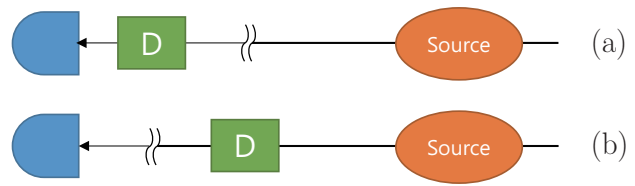}
	}
	\caption{\label{scheme1}(Color online) Schematics for (a) displaced measurement under decoherence in a part of the entangled state and (b) displaced measurement using an inefficient detector. Only one part of the Bell inequality test is shown. 
	}
\end{figure}

\begin{figure}[t]
	\Large
	\centering
	\resizebox{0.5\textwidth}{!}{
		\includegraphics{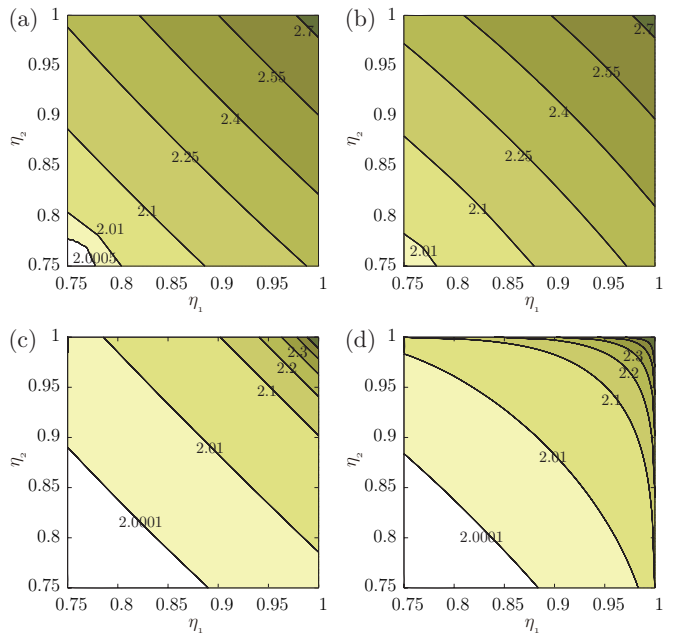}
	}
	\caption{\label{inefficient} (Color online) Numerically optimized Bell-CHSH function for all displacement variables and amplitudes using imperfect photodetectors. The detection efficiencies are $\eta_1$ for mode 1 and $\eta_2$ for mode 2. The results in (a) and (b) show the optimized Bell-CHSH functions for displaced on-off measurement using the odd ECSs. The results are restricted to the symmetric ECS in (a) while they are optimized for any values of $\alpha_1$ and $\alpha_2$ with the asymmetric ECS in (b).
The optimizing values of $\alpha_1$ and $\alpha_2$ for (a) and (b) are between $0.4$ to $1.4$.
	  Results with parity measurements are shown in (c) for the even symmetric ECS and (d) for the even asymmetric ECS with any values of $\alpha_1$ and $\alpha_2$. The two uppermost lines in (c) and (d) indicate 2.4 and 2.5, respectively. When $\eta_1=\eta_2=1$, the maximum violations of the Bell-CHSH functions are up to $\approx 2.743$ for (a) and (b) and $2.820$ for (c) and (d).
	  The optimizing values of $\alpha_1$ and $\alpha_2$ for (c) and (d) are between $0.02$ to $1.64$ for $\eta_1<0.99$ and $\eta_2<0.99$.
	  }
\end{figure}

In detail, for the case of the displaced on-off measurements with the odd ECS (Figs.~\ref{inefficient}(a) and ~\ref{inefficient}(b)), optimizing values of $\alpha_1$ and $\alpha_2$ lie between $0.4$ to $1.4$, which is experimentally feasible \cite{ourjoumtsev09,Ourj2007}. We find that the asymmetric  ECS (Fig~\ref{inefficient}(b)) shows  larger violations compared to the symmetric ECS (Fig~\ref{inefficient}(a)). 
When $\eta=\eta_1=\eta_2$, a detection efficiency of $\eta \ge 0.771$ 
is required for violation of $\abs{\mathcal{B}_O}\geq 2.001$ with the symmetric ECS
while a smaller efficiency $\eta \ge 0.745$ is sufficient for the same amount of violation with the asymmetric ECS. When the efficiencies for modes 1 and 2 are $\eta_1=0.75$  and $\eta_2=1$, the asymmetric ECS shows the optimized Bell quantity of $2.305$ which is larger than $2.269$ for the symmetric ECS.

In the case of the parity measurements using the even ECSs, the improvement by the asymmetric ECS is even larger.
For example, comparing $\eta_1=0.98$ line in Fig.~\ref{inefficient}(c) and (d), we find that there is a violation $\abs{\mathcal{B}_\Pi}\geq2.01$ for $\eta_2 \ge 0.760$ if we use the asymmetric form of ECS. However, much larger efficiency $\eta_2\ge 0.805$ is needed when we use the symmetric ECS.
The optimizing values of $\alpha_1$ and $\alpha_2$ lie between $0.02$ to $1.64$ if $\eta_1<0.99$ and $\eta_2<0.99$. We also note that if one of the detectors is perfect, the optimizing value of the amplitude of that mode goes to infinity. 
This behavior can be inferred from the results in Sec.~\ref{subsec:parity_loss} with Fig.~9(b) for the case of decoherence, noting that the decoherence and the detection inefficiency give qualitatively the same effects (Appendix B). When both the detectors are perfect, of course, the larger amplitude gives the larger violation for each mode as implied in Fig.~4; the maximum violation appears when both $\alpha_1$ and $\alpha_2$ become infinity.

\section{\label{sec:remarks}Remarks}
In this paper, we have studied asymmetric ECSs as well as asymmetric lossy environments
for Bell-CHSH inequality tests.
We first investigated the violations of Bell-CHSH inequality using perfect on-off detectors and ideal ECSs. 
We have shown that the asymmetric form of ECS could give larger violations of the Bell-CHSH inequality in some region of the averaged photon numbers of the ECSs. 
On the other hand, in the case of photon number parity measurements, we could not find apparent improvement of the Bell violations using the asymmetric form of ECS with perfect detectors.

We then studied Bell-CHSH violations under the effects of decoherence on the ECSs. We considered two different  schemes for distributing entanglement under photon loss, i.e., symmetric and asymmetric schemes. In the symmetric scheme, the photon losses occur in both modes of the ECSs. On the other hands, photon loss occurs only in one of the two modes in the case of the asymmetric scheme. We showed that the asymmetric form of ECS can increase the amount of violations significantly under the asymmetric scheme compared to the case of the symmetric scheme. For example, when the normalized time is $r=0.2$, the Bell-CHSH function using photon on-off detectors for the asymmetric loss scheme shows the maximum value of $2.145$, which is much larger than that for the symmetric case, $2.054$. A similar improvement can be made in the case of photon number parity measurements;
when $r=0.1$ with the ECS of average photon number $\approx 4.19$, the optimized Bell-CHSH function  under the symmetric loss scheme is about $2.014$  but
it is $2.131$  under the asymmetric scheme.

We have also investigated effects of inefficient detectors. We show that the asymmetric form of ECSs lowers the detection efficiency required for violation of the Bell-CHSH inequality.
For example, a detection efficiency of $\eta \ge 0.771$ 
is required for a Bell-CHSH violation of $\abs{\mathcal{B}_O}\geq 2.001$ with the symmetric ECS, but
a smaller efficiency $\eta \ge 0.745$ is sufficient for the same amount of violation with the asymmetric ECS. 
This improvement is even more significant for the case of parity measurements particularly
when the detection efficiencies of two detectors differ much. When the detection efficiency for mode 1 is  $\eta_1=0.98$, the required detection efficiency for mode 2 is $\eta_2 \gtrsim 0.81$ to show the violation of $|B_\Pi| \geq 2.01$ using the symmetric ECS is used, but it is $\eta_2 \gtrsim 0.76$ when the asymmetric ECS is used.
The optimizing amplitudes are found between $0.4$ to $1.4$ in the case of the on-off measurements and between $0.02$ to $1.64$ for the parity measurements. It is worth noting that these values of amplitudes for ECSs are within reach of current technology \cite{ourjoumtsev09,Ourj2007,sasaki-cat,nam-cat}.

In summary, our extensive study reveals that the asymmetric form of ECSs and the asymmetric scheme for distributing entanglement enable one to effectively test the Bell-CHSH inequality with the same resources.

\begin{acknowledgments}
This work was supported by the National Research Foundation of Korea (NRF) grant funded by the Korea government (MSIP) (No. 2010-0018295) and by the Center for Theoretical Physics at Seoul National University. 
The numerical calculations in this work were performed using Chundoong cluster system in the Center for Manycore Programming at Seoul National University. 
\end{acknowledgments}

\appendix

\section{\label{sec:corr_func}Correlation functions for on-off and parity measurements}
Here, we present the explicit forms of the correlation functions defined in Eqs.~\eqref{E_onoff} and \eqref{E_parity} under lossy effects. Instead of computing the correlation function for every case, we calculate the results for an ECS under the beam splitters with transmissivities $\eta_1$ and $\eta_2$ for mode 1 and mode 2, respectively, and show that these results are applicable to all the cases discussed in this paper.

The beam splitter operator with transmissivity $\sqrt{\eta}$ between modes $a$ and $b$ can be represented by  $\hat{B}_{ab} = \exp[(\cos^{-1}\sqrt{\eta})(\hat{a}_a^\dagger\hat{a}_b-\hat{a}_a\hat{a}_b^\dagger)]$ \cite{campos89}. 
If a coherent-state dyadic $\ket{\alpha}\bra{\beta}$ for mode $C$ is mixed with the vacuum for mode $v$, the initial coherent-state dyadic becomes
\begin{align*}
	&\Tr_v[\hat{B}_{Cv}(\ket{\alpha}\bra{\beta})_C\otimes(\ket{0}\bra{0})_v\hat{B}_{Cv}^\dagger] \\
	&= \exp\bigl[-\frac{1}{2}(1-\eta)(\abs{\alpha}^2+\abs{\beta}^2-2\alpha\beta^*)\bigr]\ket{\sqrt{\eta}\alpha}\bra{\sqrt{\eta}\beta}.\label{dkb}\numberthis
\end{align*}
We now apply this result to a pure ECS $\ket{\rm ECS^\pm}$ mixed with the vacuum modes through two beam splitters with  transmissivities $\eta_1$ and $\eta_2$, and the result is
\begin{align*}
	\rho^{\pm}&[\alpha_1,\alpha_2,\eta_1,\eta_2]\\
	&= \mathcal{N_\pm}^2 \bigl\{ \ket{\sqrt{\eta_1}\alpha_1}\bra{\sqrt{\eta_1}\alpha_1}\otimes\ket{\sqrt{\eta_2}\alpha_2}\bra{\sqrt{\eta_2}\alpha_2} \\
	&\pm e^{-2[(1-\eta_1)\alpha_1^2+(1-\eta_2)\alpha_2^2]}&\\ 
	&\bigl[\ket{\sqrt{\eta_1}\alpha_1}\bra{-\sqrt{\eta_1}\alpha_1}\otimes\ket{\sqrt{\eta_2}\alpha_2}\bra{-\sqrt{\eta_2}\alpha_2} \\
	&+ \ket{-\sqrt{\eta_1}\alpha_1}\bra{\sqrt{\eta_1}\alpha_1}\otimes\ket{-\sqrt{\eta_2}\alpha_2}\bra{\sqrt{\eta_2}\alpha_2} \bigr] \\
	&+ \ket{-\sqrt{\eta_1}\alpha_1}\bra{-\sqrt{\eta_1}\alpha_1}\otimes\ket{-\sqrt{\eta_2}\alpha_2}\bra{-\sqrt{\eta_2}\alpha_2} \bigr\}.
\end{align*}
It is straightforward to check that if we let $\eta_1=\eta_2=t$ then the state will be exactly the same to Eq.~\eqref{rhoA}. In addition, $\eta_1=1$ with $\eta_2=t^2$ will give Eq.~\eqref{rhoB}.

Using the result, the correlation functions for on-off measurement and parity measurement are obtained as
\begin{widetext}
\begin{align*}
	E_O(\xi,\chi) &= \Tr[\rho^\pm \hat{O}_1(\xi)\otimes\hat{O}_2(\chi)]\\
		&=\frac{2}{2 \pm 2 e^{-2 \alpha_1^2-2 \alpha_2^2}} 
	\bigl[1\mp2 e^{-2 \alpha_1^2-2 \alpha_2^2-\chi_i^2-\chi_r^2-\xi_i^2-\xi_r^2} \bigl(e^{\alpha_1^2 \eta_1 + \chi_i^2 + \chi_r^2} \cos \bigl(2 \alpha_1 \sqrt{\eta_1} \xi_i\bigr) \\
		&+e^{\alpha_2^2 \eta_2+\xi_i^2+\xi_r^2} \cos \bigl(2 \alpha_2 \chi_i \sqrt{\eta_2}\bigr) \bigr) -2 e^{\alpha_1^2 \eta_1 + \alpha_2^2 \eta_2} \cos \bigl(2 \bigl(\alpha_1 \sqrt{\eta_1} \xi_i+ \alpha_2 \sqrt{\eta_2}\chi_i \bigr)\bigr)\\
	& \pm e^{-2 \bigl(\alpha_1^2+\alpha_2^2\bigr)} +2 e^{-\bigl(\xi_r-\alpha_1 \sqrt{\eta_1}\bigr)^2-\bigl(\chi_r-\alpha_2 \sqrt{\eta_2}\bigr)^2-\chi_i^2-\xi_i^2} +2 e^{-\bigl(\alpha_1 \sqrt{\eta_1}+\xi_r\bigr)^2-\bigl(\alpha_2 \sqrt{\eta_2}+\chi_r\bigr)^2-\chi_i^2-\xi_i^2}\\
	&-e^{-\bigl(\xi_r - \alpha_1 \sqrt{\eta_1}\bigr)^2-\xi_i^2}-e^{-\bigl(\alpha_1 \sqrt{\eta_1}+\xi_r\bigr)^2-\xi_i^2}
-e^{-\bigl(\alpha_2 \sqrt{\eta_2}+\chi_r\bigr)^2-\chi_i^2}-e^{-\bigl(\chi_r-\alpha_2 \sqrt{\eta_2}\bigr)^2-\chi_i^2}\bigr], \\
E_\Pi(\xi,\chi) &= \Tr[\rho^\pm \hat{\Pi}_1(\xi)\otimes\hat{\Pi}_2(\chi)]\\
	&=\frac{1}{2\pm 2e^{-2 \alpha_1^2-2 \alpha_2^2}}\exp \left(-2 \left(\alpha_1^2 (\eta_1+1) + \alpha_2^2 (\eta_2+1)+\chi_i^2+\chi_r^2+\xi_i^2+\xi_r^2\right)\right)\\
&\left(\pm2e^{4 \left(\alpha_1^2 \eta_1 + \alpha_2^2 \eta_2\right)} \cos \left(4 \alpha_1 \sqrt{\eta_1} \xi_i+4 \alpha_2 \chi_i \sqrt{\eta_2}\right)+e^{2 \left(\alpha_1^2 -2\alpha_1\xi_r\sqrt{\eta_1} + \alpha_2^2 -2 \alpha_2 \chi_r \sqrt{\eta_2}\right)} \left(e^{8 \alpha_1 \sqrt{\eta_1} \xi_r+8 \alpha_2 \chi_r \sqrt{\eta_2}}+1\right)\right),
\end{align*}
\end{widetext}
where $\xi=\xi_r+i\xi_i$ and $\chi=\chi_r+i\chi_i$, and $O_i(\xi)$ and $\Pi_i(\xi)$ were defined in Eqs.~\eqref{eq:O} and \eqref{eq:Pi}. By substituting $\eta_1=\eta_2=1$, we can get a correlation functions for the case of perfect detectors. Likewise, $\eta_1=\eta_2=t$ gives the correlation functions for distribution strategy A and $\eta_1=1$, $\eta_2=t^2$ gives the correlation function for strategy B. As we will discuss below, we also can use these results for the cases of imperfect detectors even though they differ in the order of the displacement operators and the beam splitters.

\section{\label{sec:equiv}The order of photon loss and displacement operator in correlation}
In this section, we show that ECSs give the same optimized value of the Bell-CHSH function independent to the order of the displacement and the photon loss by a beam splitter. Note that it is sufficient to only consider two coherent-state dyadics of the form $\ket{\gamma}\bra{\gamma}$ and $\ket{\gamma}\bra{-\gamma}$. This is because ECSs only contain this kind of dyadics and a photon loss by a beam splitter is superlinear(linear in the space of density matrices) and a displacement operator is linear.

First, suppose that the states undergo a beam splitter first and displacement second. From the above, the dyadics after the transmission to the beam splitter with transmissivity $\sqrt{\eta}$ would be
\begin{align*}
	\ket{\gamma}\bra{\gamma} &\longrightarrow \ket{\sqrt{\eta}\gamma}\bra{\sqrt{\eta}\gamma} \\
	\ket{\gamma}\bra{-\gamma} &\longrightarrow \exp[-2(1-\eta)\abs{\gamma}^2]\ket{\sqrt{\eta}\gamma}\bra{-\sqrt{\eta}\gamma}.
\end{align*}
Now we apply the displacement operator $\hat{D}(\xi) = \exp[\xi\hat{a}^\dagger-\xi^*\hat{a}]$ to photon lost dyadics. Then the states would be
\begin{align*}
	\ket{\gamma}\bra{\gamma} \longrightarrow& \ket{\sqrt{\eta}\gamma-\xi}\bra{\sqrt{\eta}\gamma-\xi} \\
	\ket{\gamma}\bra{-\gamma} \longrightarrow& \exp[-2(1-\eta)\abs{\gamma}^2+\sqrt{\eta}(\xi^*\gamma-\xi\gamma^*)]\\
	&\quad\ket{\sqrt{\eta}\gamma-\xi}\bra{-\sqrt{\eta}\gamma-\xi}.\numberthis \label{disloss}
\end{align*}

Next, let us consider the case we employ the operators to opposite order, displacement first and photon loss second. After the displacement, the dyadics would be
\begin{align*}
	\ket{\gamma}\bra{\gamma} \longrightarrow& \ket{\gamma-\xi}\bra{\gamma-\xi} \\
	\ket{\gamma}\bra{-\gamma} \longrightarrow& \exp[\xi^*\gamma-\xi\gamma^*]\ket{\gamma-\xi}\bra{-\gamma-\xi}.
\end{align*}
Applying photon loss to these dyadics using Eq.~\eqref{dkb}, we finally obtain
\begin{align*}
	\ket{\gamma}\bra{\gamma} \longrightarrow& \ket{\sqrt{\eta}(\gamma-\xi)}\bra{\sqrt{\eta}(\gamma-\xi)} \\
	\ket{\gamma}\bra{-\gamma} \longrightarrow& \exp[-2(1-\eta)\abs{\gamma}^2+\eta(\xi^*\gamma-\xi\gamma^*)]\\
		&\quad\ket{\sqrt{\eta}(\gamma-\xi)}\bra{\sqrt{\eta}(-\gamma-\xi)}.\numberthis \label{lossdis}
\end{align*}
These results show that if we replace $\xi$  with $\sqrt{\eta}\xi$  in Eq.~\eqref{disloss}, it will be exactly the same to Eq.~\eqref{lossdis}. This means that the optimized values of the Bell-CHSH functions for all displacement variables do not depend on the order of loss and displacement.


\begin{thebibliography}{99}
\bibitem{note-num}
One may ask whether this unsmooth change implies that our results around  $\bar{n}_-\approx1.2$
are not the global maxima. 
We cannot exclude such a possibility because we do not have an analytical evidence to claim that all the results
obtained using the  BFGS algorithm \cite{fletcher87}  throughout the paper are the global maxima.
This is a typical problem in an optimization process with multiple parameters as we do in the present paper with 8 to 10 parameters (assuming $\alpha_1$ and $\alpha_2$ are real without loss of generality).
\end{thebibliography}
\end{document}